\documentclass[runningheads]{llncs}
\usepackage{graphicx}
\usepackage{listings}
\usepackage{xcolor}
\usepackage{amsmath}
\usepackage{bbm}
\usepackage{dsfont}
\usepackage{lipsum}
\usepackage[most]{tcolorbox}
\usepackage[colorlinks = true,
            linkcolor = blue,
            urlcolor  = blue,
            citecolor = blue,
            anchorcolor = blue]{hyperref}

\newtcolorbox{shadedcvbox}[1][]{enhanced jigsaw,
  colback=white!80!blue,
  coltext={black},
  boxrule=0pt,
  arc=3mm,
  auto outer arc,
  boxsep=3pt,
  left=4pt,
  right=2pt,
  bottom=2pt,
  top=2pt,
  fontupper={\bfseries},
  #1}

\definecolor{codegreen}{rgb}{0,0.6,0}
\definecolor{codegray}{rgb}{0.5,0.5,0.5}
\definecolor{codepurple}{rgb}{0.58,0,0.82}
\definecolor{backcolour}{rgb}{0.95,0.95,0.92}

\lstdefinestyle{mystyle}{
  backgroundcolor=\color{backcolour}, commentstyle=\color{codegreen},
  keywordstyle=\color{magenta},
  numberstyle=\tiny\color{codegray},
  stringstyle=\color{codepurple},
  basicstyle=\ttfamily\footnotesize,
  breakatwhitespace=false,         
  breaklines=true,                 
  captionpos=b,                    
  keepspaces=true,                 
  numbers=left,                    
  numbersep=5pt,                  
  showspaces=false,                
  showstringspaces=false,
  showtabs=false,                  
  tabsize=2
}

\begin{document}
\title{Usable Region Estimate for Assessing Practical Usability of Medical Image Segmentation Models}
%
%\titlerunning{Abbreviated paper title}
% If the paper title is too long for the running head, you can set
% an abbreviated paper title here
%
%\author{***}

%\institute{***}
\author{Yizhe Zhang\inst{1} \and
Suraj Mishra\inst{2} \and
Peixian Liang\inst{2}\and
Hao Zheng\inst{2}\and
Danny Z. Chen\inst{2}
}
%index{Zhang, Yizhe} 
%index{Mishra, Suraj} 
%index{Liang, Peixian} 
%index{Zheng, Hao} 
%index{Chen, Danny Z.} 
%
%\authorrunning{F. Author et al.}
% First names are abbreviated in the running head.
% If there are more than two authors, 'et al.' is used.
%
\institute{School of Computer Science and Engineering, Nanjing University of Science and Technology, Nanjing, Jiangsu 210094, China\\ \email{yizhe.zhang.cs@gmail.com}\and Department of Computer Science and Engineering, University of Notre Dame,
Notre Dame, IN 46556, USA}
%Springer Heidelberg, Tiergartenstr. 17, 69121 Heidelberg, Germany
%\email{lncs@springer.com}\\
%\url{http://www.springer.com/gp/computer-science/lncs} \and
%ABC Institute, Rupert-Karls-University Heidelberg, Heidelberg, Germany\\
%\email{\{abc,lncs\}@uni-heidelberg.de}}
%
\maketitle              % typeset the header of the contribution
\begin{abstract}
We aim to quantitatively measure the practical usability of medical image segmentation models: to what extent, how often, and on which samples a model's predictions can be used/trusted. We first propose a measure, Correctness-Confidence Rank Correlation (CCRC), to capture how predictions' confidence estimates correlate with their correctness scores in rank. A model with a high value of CCRC means its prediction confidences reliably suggest which samples' predictions are more likely to be correct. Since CCRC does not capture the actual prediction correctness, it alone is insufficient to indicate whether a prediction model is both accurate and reliable to use in practice. Therefore, we further propose another method, Usable Region Estimate (URE), which simultaneously quantifies predictions' correctness and reliability of confidence assessments in one estimate. URE provides concrete information on to what extent a model's predictions are usable. In addition, the sizes of usable regions (UR) can be utilized to compare models: A model with a larger UR can be taken as a more usable and hence better model. Experiments on six datasets validate that the proposed evaluation methods perform well, providing a concrete and concise measure for the practical usability of medical image segmentation models. Code is made available at \href{https://github.com/yizhezhang2000/ure}{https://github.com/yizhezhang2000/ure}.%We also highlight several exciting observations in experiments. 

%For real-world clinical use cases, medical AI models meet challenges in deployment because of reliability and trustworthiness issues. If we do not design our evaluation methods to address these issues, model development would not fully address them. 

%Fairness in applying machine learning methods for medical imaging application is a key xxxx. Previous work has been addressing fairness among xxx, mostly in the model building and training perspectives. In this paper, we focus on fairness in evaluating machine learning models. When comparing the testing performance between model A and model B, it is commonly to use the performance averaged over all in instances/individuals. We argue that in the early stage of Medical AI development, this metric is fine but at the current stage, when improvement become sutble, and fairness become more and more important in clinical practice, we need better, more precise and more fair evaluation protocol. We propose InsCompare: a new evaluation protocol to evaluate/compare two models on the level of instance/individuals. Namely, the performance is compared on the instance level, and overall comparison is done based on a distribution of performance difference rather a single number. We use the xxx, xxx, xxx models on xxx, xxx, xxx datasets to demonstrate xxx.

%The abstract should briefly summarize the contents of the paper in
%15--250 words.
\end{abstract}
\section{Introduction}
\begin{figure}
    \begin{quote}
    \begin{shadedcvbox}
        Conventional Eval.: ``This model achieves 0.87 ($\pm$0.18) $F_1$ score and 0.19 ECE (Expected Calibration Error) on the test data.''
    \end{shadedcvbox}
    \begin{shadedcvbox}[colback=gray!20!white,coltext={black},drop fuzzy shadow={gray!80!black}]
%\vspace{-0.05in}
    Doctor: ``OK. If future samples are from the same/similar distribution of these test samples, then how often and to what extent the model's predictions are usable in practice?''
    \end{shadedcvbox}   
%\vspace{-0.05in}
%\begin{center}
%  $\diamondsuit$~$\diamondsuit$~$\diamondsuit$
%\end{center}
     \begin{shadedcvbox}
        URE (proposed).: ``Predictions with confidences no smaller than 0.76 satisfy the pre-defined clinical correctness requirement, and 83\% of the time, predictions are in this usable region.''
                \end{shadedcvbox}
    \begin{shadedcvbox}[colback=gray!20!white,coltext={black},drop fuzzy shadow={gray!80!black}]

        Doctor: ``Good! I will use this model following the 0.76 confidence guideline.'' (OR: ``Not good enough. I would like the predictions being usable 95\% of the time. Please continue to improve.'')
             \end{shadedcvbox}  
     \end{quote}\label{fig.1:dialog}
        \caption[]{A dialog example to show the merit of clinically oriented model evaluations. Our estimate simultaneously evaluates both predictions' correctness and confidence estimates of a given prediction model, providing a quantitative measure for how usable the model is and concrete guidelines for using the model.}

\end{figure}

%Conventional eval.: Model A achieves $x$ $F_1$ score, with ECE for confidence estimations measured as $y$ for the test set.\\
%Doctors: OK, how reliable are the predictions? If the future samples drawn from the same distribution of the testing samples, how much and to what extent the model's predictions can be trusted?

Deep learning (DL) methodologies brought performances of medical image segmentation, detection, and classification to new heights. Many new medical AI prediction models achieved impressive accuracy (correctness) on test sets. In the meantime, practitioners still face reliability/trustworthiness issues when deploying AI models to real clinical scenarios. Conventional model evaluation measures (e.g., average $F_1$ score, average Dice Coefficient) mainly focus on prediction correctness. Although a model with good correctness scores is desirable, high correctness alone is not sufficient for being highly usable in clinical practice. As each model inevitably would make some wrong predictions, a highly usable/reliable model should also have its prediction confidence estimates well aligned with the prediction correctness so that the model can alert experts for further input when its confidences for some samples' predictions are low.

Improving the calibration and reliability of confidence estimates is an active research area. Guo et al.~\cite{guo2017calibration} found that a DL model is often overconfident in its predictions; they examined a range of methods to calibrate model output confidences. For example, temperature scaling gives the best empirical calibration results due to the strong non-linearity nature; it requires additional labeled data for calibration and does not change the rank order of the original uncalibrated confidence estimates. In \cite{lakshminarayanan2016simple,mehrtash2020confidence}, it was proposed to use model ensembles to improve the reliability of prediction confidence estimates. Dropout-based and Bayesian-based methods \cite{gal2016dropout,maddox2019simple} are also popular choices for better uncertainty/confidence estimates. On the evaluations side, Expected Calibration Error (ECE) \cite{guo2017calibration,mehrtash2020confidence,popordanoska2021relationship} is widely used for evaluating calibration of prediction confidences. ECE computes the absolute difference between the correctness score and confidence score for each prediction (or each bin of predictions) and takes the average of the computed absolute differences as output. Low ECE values indicate that confidences are better matched with the actual correctness scores. Similarly, Brier score (Br)~\cite{brier1950verification} measures the mean square difference between each probabilistic prediction and its corresponding ground truth label. ECE and Br quantify how well a model's confidence estimates (CE) match with correctness scores (CS) in value but do not measure alignment/matching in rank. 

%CE that aligned well with CS in rank is informative to indicate which prediction would be more likely to be correct. Interestingly, a good score in a rank-based correlation between CE and CS does not necessarily yield high recommendation in E/M/B scores since any level of global shifts in values of CE would lead to worse E/M/B scores but may not affect a rank correlation between CE and CS. On the other hand, small changes in values may significantly affect correctness-confidence alignment in rank but may not be pronounced in E/M/B scores. Therefore, for evaluating the reliability of CE, we believe a rank correlation between CE and CS would bring more insights, besides using the ECE type of measure alone.

More clinically oriented model evaluations drive more clinically oriented model developments. Prediction correctness (e.g., $F_1$ score, Dice coefficient) and goodness of alignment between confidence estimates and correctness scores capture two key aspects of a clinically usable medical AI model. But, neither correctness scores nor confidence estimation reliability scores alone provide a unified view on practical usability of a model. For example, a 0.87 average $F_1$ score and 0.19 ECE score inform little on how one could utilize such a model in practice, and a 0.02 improvement in $F_1$ score (sample-average) and $0.01$ increase in ECE do not offer concrete ideas on how it affects the model's practical usability. There is a strong need to develop a clinically oriented model evaluation method that can inform users to what extent the model's predictions are usable and how often the model's predictions can be used in practice (see Fig.~\ref{fig.1:dialog}).

\begin{figure}[t]
\includegraphics[width=\textwidth]{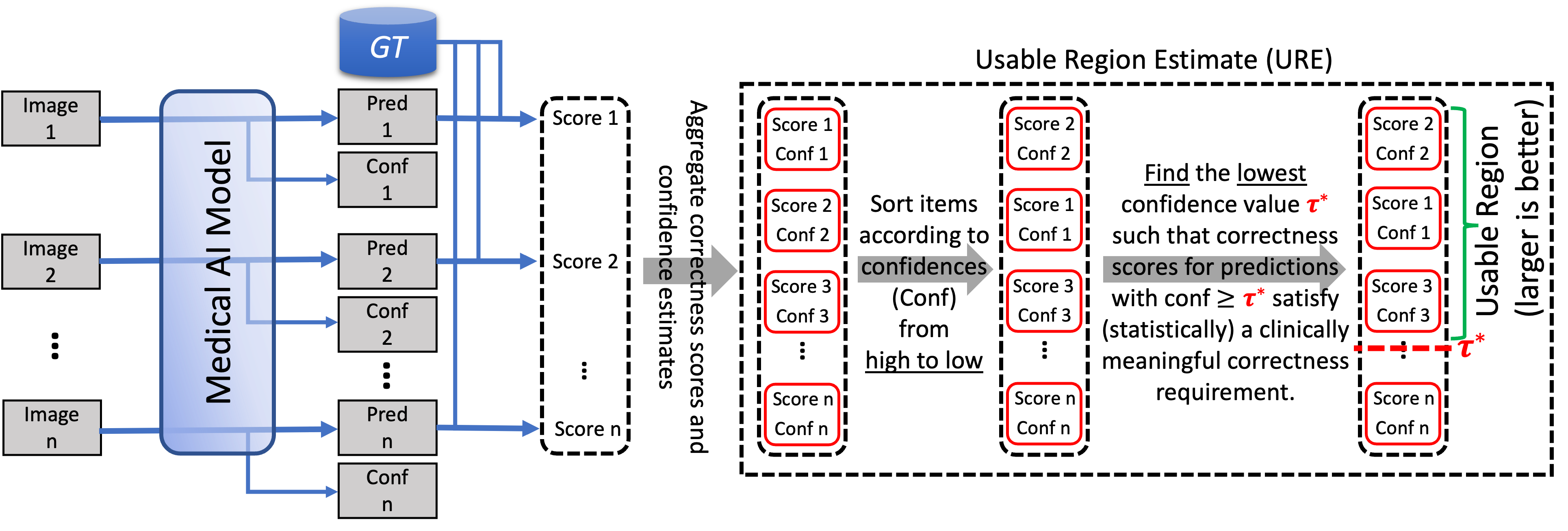}
\caption{Illustrating the proposed Usable Region Estimate (URE).}
\label{fig2:highlevel}
\end{figure}

In this paper, we first propose a new measure, Correctness-Confidence Rank Correlation (CCRC), to assess the goodness of alignment between correctness scores and confidence estimates from the rank perspective. A high CCRC value indicates a sample's prediction with a {higher} confidence is more likely to be correct (and vice versa). Furthermore, we design a novel estimate named Usable Region Estimate (URE), which quantifies prediction correctness and reliability of confidence estimates in one statistically sound and unified measure, where only \textbf{high correctness scores} together with \textbf{properly ordered confidence estimates} would yield good evaluation results (see Fig.~\ref{fig2:highlevel}). URE measures to what extent in the confidence space the predictions are usable (with high quality), giving a concrete guideline for using the model in practice and allowing direct comparison between models via comparing the sizes of usable regions.

Our work contributes to clinically oriented and more comprehensive model evaluations in the following aspects. (1) We identify limitations of previous confidence evaluation measures (e.g., ECE), and propose CCRC that takes rank information into account in measuring reliability of confidence estimates. (2) We identify that known methods which separately evaluate prediction correctness and reliability of confidence estimates do not offer a unified measure for a model's practical usability. (3) We propose a novel statistically sound Usable Region Estimate that unifies evaluations of prediction correctness and reliability of confidence estimates in one measure. URE can quantitatively measure a model's usability and give concrete recommendations on how to utilize the model.

%\subsection{Motivation and High-level Ideas}
%A model's prediction should be accurate enough, meanwhile its confidences associated with the predictions should also be informative in determining which prediction is reliable and which prediction requires human intervention.
\section{Methodology}
\subsection{Preliminaries}
\noindent \textbf{Correctness of Predictions.} Given predictions ${\hat{y_1}, \hat{y_2}, \ldots, \hat{y_n}}$ and ground truth labels ${{y_1}, {y_2}, \ldots, {y_n}}$, a correctness metric $\phi$ (e.g., $F_1$ score) applies to each pair of prediction and ground truth label to obtain a correctness score $s_i=\phi(\hat{y}_i, y_i)$. An overall correctness score is defined by $\sum_{i=1}^{n} \frac{s_i}{n}$. 

%A general form can be written as $\sum_{i=1}^{n} \frac{w_i \times s_i}{\sum_{i=1}^{n}{w_i}}$, where $w_1, w_2, \dots, w_n$ are sample-specific importance weights.

\noindent \textbf{Reliability of Confidence Estimates.} \underline{NLL:} Given an medical AI model $\hat{\pi}(Y|X)$ and $n$ test samples, NLL is defined as $L_{NLL}=\sum_{i=1}^{n}-\log(\hat{\pi}(y_i|x_i))$. When $\hat{\pi}(y_i|x_i)$ is identical to the ground truth conditional distribution $\pi(Y|X)$, the value of $L_{NLL}$ is minimized. Since an overly confident model tends to give a low NLL value, the NLL value is not a reliable indicator for the goodness of alignment between confidence estimates and predictions' correctness. \underline{ECE:} For a given test set of $n$ samples, one obtains the model confidence of each sample, $conf_i$, and the accuracy/correctness score $s_i$, for $i = 1, 2, \dots, n$. One can compare them at the per-sample level or the bin level. For per-sample level comparison, $ECE = \sum_{i=1}^{n} \frac{1}{n} |s_i - conf_i|$. \underline{Brier score (Br):} Suppose a model prediction ${\hat{y_i}}$ is in a probabilistic form; Br measures the distance between the prediction and ground truth: $Br = {\sum_{i=1}^{n}\frac{1}{n}(\hat{y_i}-y_i)^2}$.

\subsection{Correctness-Confidence Rank Correlation (CCRC)}
Given a correctness score $s_i$ and a model confidence ${conf}_i$ for each test sample, $i=1, 2, \ldots, n$, we use Spearman's rank correlation coefficient~\cite{schober2018correlation} to measure the correlation between these two lists of numbers. The coefficient value obtained is between 0 and 1 (inclusive), with 1 for a perfect alignment in rank between confidence estimates and correctness scores. A high CCRC is critical for model deployment as the confidence estimate is informative on which sample's prediction is \textit{more} likely to be correct. Being high in CCRC is a necessary condition for a medical AI model for being highly usable in practice.%, but a high CCRC alone is not enough to conclude high usability as actual correctness of predictions is not measured by CCRC.

\subsection{Usable Region Estimate (URE)}
Given a correctness score $s_i$ and a model confidence ${conf}_i$ for each test sample, $i=1, 2, \ldots, n$, we compute the mean of the correctness scores, $\mu_{\tau}$, for all the predictions whose confidences are no smaller than a confidence level $\tau$,
$count_{{\tau}} = \sum_{i=1}^{n} \mathbbm{1}_{{\{conf_i \geq \tau\}}}$, and $\mu_{\tau} =\sum_{i=1}^{n}  \frac{s_i \mathbbm{1}_{{\{conf_i \geq \tau\}}} }{count_\tau}$.
%An absolutely fair model (GC is close to 0) is challenging to achieve in practice. Building a reliable system that gives users confidence for each prediction enables efficient and effective doctor's intervention to ensure correctness and fairness.

% we propose an evaluation method for quanty both accuracy/correctness and confidences in a concrete manner that the resulted estimate can be directly used as a guideline in clinical practice. The high-level idea is to sort confidences from high to low, and from the top to the bottom of this list (prediction with the highest confidence to the lowest confidence), we check if the overall prediction correctness is good enough for clinical practice. The more predictions in this ``reliable region'', the better the model is. 

URE aims to find the lowest possible value $\tau^*$ such that for all the predictions with confidence estimates no smaller than $\tau^*$, the lower end of the $95\%$ CI (confidence interval) of $\mu_{\tau}$ is no less than a correctness score requirement (clinically usable segmentation accuracy). We utilize the classic bootstrapping techniques \cite{efron1994introduction} for computing CI, as bootstrapping does not assume the data being in a particular distribution (e.g., normal distribution). The obtained $\tau^*$ gives a practical guideline that, using the tested model, any prediction with a confidence value $\geq \tau^*$ is considered being usable, and predictions with confidences lower than $\tau^*$ need human attention\footnote{Assume the test set for computing UR is representative of the true data distribution.}. To compute the size of the usable region of predictions, we further compute $p^*$, which is the ratio between the number of predictions with confidences $\geq\tau^*$ and the total number of predictions. A model with a larger value of $p^*$ means that this model has a larger \textbf{usable region} of predictions and is considered to be a better model compared to one with a smaller value of $p^*$. One can efficiently find $\tau^*$ by sorting the confidences and iterating through the sorted list and checking the correctness scores. The Python code for URE is given in Listing 1.1. A usability diagram can be created by computing UR using a range of correctness requirements.

%Bootstrapping 
%If the correctness constraint check is

%given a set of predictions, their associated confidences, and a clinically defined correctness constraint $t$. 

%Note that $c^*$
%When comparing two models' $c^*_{model A}$ and $c^*_{model B}$, because the xxx confidence level can differ, that is one model xxx. xxx.

%Given a model's predictions and confidence associated with each prediction, we run the above code to find $\tau^*$ and $p^*$. We then consider this model being $\tau^*$ reliable and $p^*$-th percentile reliable. 

\begin{lstlisting}[language=Python, caption= Usable Region Estimate (URE): Finding the ``Usable Region'' in which predictions with confidences $\geq\tau^*$ satisfy a clinically meaningful correctness constraint.]
def URE(scores,conf, requirement=0.9):
    index = np.argsort(-conf); tau_s=1.0; p_s=0
    for id in range(len(index)):
        pool=scores[index[:id+1]]; statistics=[];
        for b in range(1,100): # Bootstrapping for CI
            poolB=np.random.choice(pool, len(pool));
            stat=np.mean(poolB); # can be of other statistics
            statistics.append(stat);
        ordered = np.sort(statistics)
        lower = np.percentile(ordered, 2.5)
        if lower>=requirement:
            tau_s=conf[index[id]];p_s=len(pool)/len(scores);
    return tau_s, p_s
\end{lstlisting}

Below we give analyses and technical discussions for the proposed URE.

%\textbf{Causes for a Small UR.} Common conditions for yielding a small usable region (with a low value of $p^*$) include: (1) the correctness scores of predictions are mostly lower than the pre-defined correctness constraint; (2) a decent amount of predictions has correctness scores higher than the pre-defined correctness constraint (e.g., Dice $\geq$ 0.9), \textbf{but} the CE does not align well with the CS in rank order (i.e., many wrong predictions have high confidences); (3) predictions are not accurate enough, and the CE is not aligned well with the CS. 

\textbf{Clinically Usable Segmentation Accuracy (CUSA) and UR Diagrams.} Setting the CUSA is crucial for developing medical AI models (for determining whether a model is good enough). CUSA is task-dependent and is expected to be determined by medical experts. Moreover, practitioners can use a range of accuracy levels to generate UR diagrams (see Fig.~\ref{fig:EXP1}) for model evaluation. A model may perform better on a portion of samples, but worse on another portion of samples, and UR diagrams can capture such phenomena when comparing models. Mixed results in UR diagrams can be viewed as the superiority of a model over other models is unstable (unconvincing). Today, segmentation accuracy improvement is often small in the average correctness score. A model with a better average score may perform worse on a significant portion of test samples. Using the UR diagram, a model is considered superior only when it consistently gives higher numbers in URE across a range of accuracy levels. This helps avoid drawing premature decisions on which model is superior and helps the Medical AI field develop better models in a more rigorous way.

\textbf{Statistics of Interest.} For the general purpose, ``mean'' statistics was set as default to estimate the expectation of the correctness scores. For a specific application with a vital requirement on prediction correctness, one may choose to use ``2 percentile'' or ``5 percentile'' instead of ``mean'' to compute the correctness scores' statistics (in line 7 of the code snippet in Listing 1.1). %In principle, choosing which statistics or combination of statistics is application-dependent and should be decided by medical/AI practitioners and experts.

\textbf{Stability across Accuracy Levels.}
For a test set containing fewer samples, large changes on URs may occur across different accuracy levels. A larger test set leads to smoother UR changes. The smoothness of change also depends on the segmentation model itself: if there is a sharp drop in the correctness scores across two accuracy levels, the UR diagram will reflect it. Overall, using a large enough test set (e.g., $>1000$ samples), the change of UR should be smooth between different accuracy levels. This is guaranteed by the bootstrapping technique used in the URE algorithm and the central limit theorem.

\textbf{The Role of URE.} URE evaluates prediction correctness and confidence estimates simultaneously in a unified measure. It is fundamentally different from \textit{individually} using average correctness scores and CCRC/ECE for model evaluation. For example, when comparing two models, suppose one gives a higher average correctness score but a worse CCRC/ECE result and the other gives a lower average correctness score but a better CCRC/ECE result. It is then \textit{inconclusive} which model can be taken as a more usable one since correctness scores and CCRC/ECE evaluate two key aspects of a model's performance which are both critical for the model usability in practice. In contrast, the proposed URE, an estimate that simultaneously addresses prediction correctness and reliability of confidence estimates, allows to compare the two models in one measure and the one yields a higher value of $p^*$ (with a larger usable region of predictions) should be considered as a more practically usable model.

\textbf{Other Implications.} New research \cite{wu2021medical,rajpurkar2022ai} suggested that correctness of predictions and reliability of confidences should be both included in evaluations of medical AI models. URE fits this emerging need well. More accurate predictions with more reliable confidence estimates yield larger UR. Thus using URE will help encourage development of accurate and well-calibrated deep learning models. Furthermore, a better-constructed test set would lead to a more accurate UR estimate. Using URE would encourage practitioners to construct a more representative test set, which can benefit clinical practice.

\section{Experiments}
Our experiments use six public medical image segmentation datasets: (1) BUSI for breast ultrasound image segmentation~\cite{al2020dataset}; (2) ISIC for skin lesion segmentation~\cite{codella2018skin}; (3) GLAS for gland segmentation in histological images~\cite{sirinukunwattana2017gland}; (4) BCSS for breast cancer semantic segmentation in histological images~\cite{amgad2019structured}. (5) MoNuSeg for cell segmentation in histological images~\cite{kumar2019multi}. (6) PhC-C2DL-PSC~\cite{mavska2014benchmark} for cell segmentation in videos. U-Net~\cite{ronneberger2015u}, DCN~\cite{chen2016deep}, DCAN~\cite{chen2016dcan}, and TransUNet~\cite{chen2021transunet} are the models tested in the experiments. A per-sample correctness score is computed for each test sample using the metric specified for every dataset (e.g., $F_1$ score). Computing per-sample confidence estimation follows a common practice, which takes the max value of prediction logits (after softmax) for each pixel and then takes the average confidence across all the pixels for each test sample. For fair and straightforward comparisons among the networks being evaluated here, we rely on the confidence estimates from the network itself without consulting any external algorithms or models.

\subsection{CCRC and URE Provide New Insights for Model Evaluation}

\begin{figure}[t]
\includegraphics[width=\textwidth]{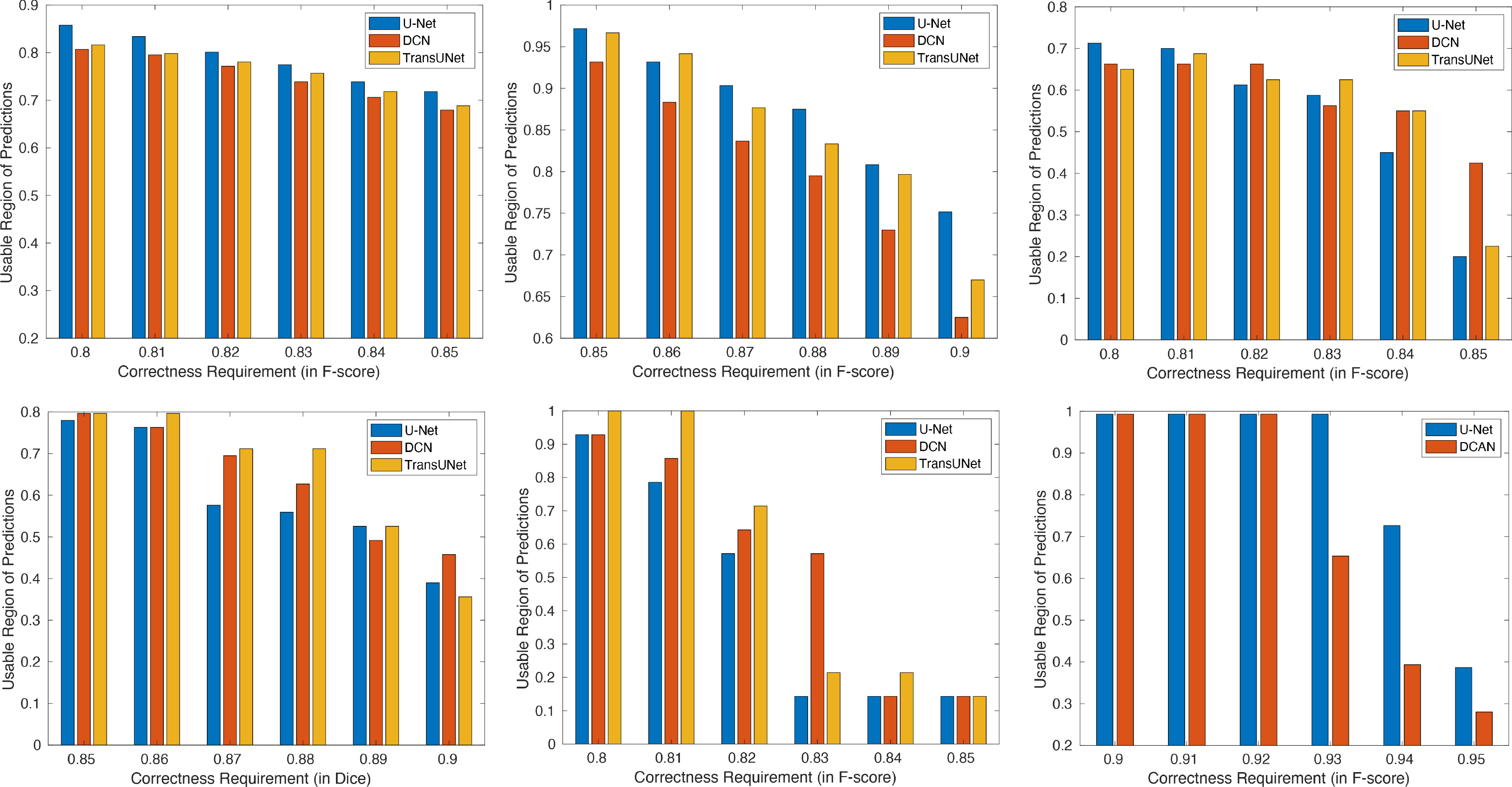}
\caption{Usability diagram: the sizes of usable regions for a range of different correctness requirements. Top-Left: BUSI~\cite{al2020dataset}. Top-Mid: ISIC~\cite{codella2018skin}. Top-Right: GLAS~\cite{sirinukunwattana2017gland}. Bottom-Left: BCSS~\cite{amgad2019structured}. Bottom-Mid: MoNuSeg~\cite{kumar2019multi}. Bottom-Right: PhC-C2DL-PSC~\cite{mavska2014benchmark}.}
\label{fig:EXP1}
\end{figure} 

\textbf{BUSI Dataset.} U-Net, DCN, and TransUNet achieve 0.721, 0.690, and 0.716 in $F_1$ score ($\uparrow$), 0.113, 0.116, and 0.113 in ECE ($\downarrow$), and 0.854, 0.865, and 0.867 in CCRC ($\uparrow$), respectively. According to $F_1$ score, {U-Net performs better than TransUNet with a small margin}. ECE suggests that TransUNet and U-Net give the same level of calibration on confidence estimates, but CCRC suggests that TransUNet's confidence estimates are considerably better aligned with the actual correctness scores. We show the sizes of usable regions estimated by the proposed URE for these networks in Fig.~\ref{fig:EXP1}(Top-Left). {U-Net provides larger usable regions} across all the correctness requirement settings.

\noindent \textbf{ISIC Dataset.} U-Net, DCN, and TransUNet achieve 0.849, 0.834, and 0.853 in $F_1$ score ($\uparrow$), 0.073, 0.094, and 0.082 in ECE ($\downarrow$), and 0.731, 0.709, and 0.659 ($\uparrow$) in CCRC, respectively. According to $F_1$ score, {TransUnet performs the best}. However, ECE and CCRC both suggest that U-Net performs the best on the reliability of confidence estimates. In Fig.~\ref{fig:EXP1}(Top-Mid), we observe that U-Net is the most usable model on ISIC for most the correctness requirement settings. 

\noindent 
%\textcolor{red}{
\noindent \textbf{GLAS Dataset.} U-Net, DCN, and TransUNet achieve 0.747, 0.773, and 0.758 in $F_1$ score ($\uparrow$), 0.214, 0.183, and 0.197 in ECE ($\downarrow$), and 0.622, 0.683, and 0.641 ($\uparrow$) in CCRC, respectively. {DCN gives better average $F_1$ score, ECE score, and CCRC score.} In Fig.~\ref{fig:EXP1}(Top-Right), URE suggests that U-Net provides the largest UR under a lower correctness requirement (0.8 in $F_1$ score). For a higher correctness requirement (0.85 in $F_1$ score), {DCN yields larger usable regions} with its predictions. TransUNet performs considerably better than U-Net for higher requirements on correctness (0.82 in $F_1$ score or higher).
%}

\noindent 
%\textcolor{red}{
\noindent \textbf{BCSS Dataset.} U-Net, DCN, and TransUNet achieve 0.832, 0.824, and 0.835 in Dice coefficient ($\uparrow$), 0.053, 0.064, and 0.063 in ECE ($\downarrow$), and 0.751, 0.809, and 0.700 ($\uparrow$) in CCRC, respectively. Comparing to U-Net, TransUNet gives slightly better Dice coefficients but worse ECE and CCRC scores. In Fig.~\ref{fig:EXP1}(Bottom-Left), URE suggests that TransUNet provides the largest UR of predictions for most the correctness requirement settings.
%}

\noindent{\textbf{MoNuSeg Dataset.} UNet, DCN, and TransUNet attain 0.819, 0.803, and 0.828 in $F_1$ score ($\uparrow$), 0.074, 0.075, and 0.072 in ECE ($\downarrow$), and 0.718, 0.815, and 0.635 ($\uparrow$) in CCRC, respectively. TransUNet performs better in the average correctness ($F_1$ score), but worse in confidence estimates (ECE and CCRC). Simultaneously measuring these two aspects, URE suggests that TransUNet is a more usable model for most the correctness requirements (see Fig.~\ref{fig:EXP1}(Bottom-Mid)).

\noindent \textbf{PhC-C2DL-PSC Dataset.} U-Net and DCAN achieve 0.937 and 0.924 in $F_1$ score ($\uparrow$), 0.023 and 0.038 in ECE ($\downarrow$), and 0.701 and 0.754 in CCRC ($\uparrow$), respectively. {U-Net gives a better average $F_1$ score, better ECE score, but worse CCRC score.} In Fig.~\ref{fig:EXP1}(Bottom-Right), we show the sizes of usable regions for these two networks. Both the networks give nearly perfect usable regions when the correctness requirement is lower. For a higher correctness requirement, {U-Net yields considerably larger usable regions} with its predictions.

URE evaluates the prediction correctness and reliability of confidence estimates simultaneously in a unified measure. As validated above, URE provides performance evaluation insights and information that conventional overall correctness scores and confidence calibration errors cannot fully capture.

\begin{table*}[t]
\footnotesize
\begin{center}
\caption{Testing the robustness of the estimated UR: the frequency ($\%$) of new samples violating the correctness requirement associated with the UR (the lower the better).}
\begin{tabular}{| c | c | c |  c |  c |  c| c| } 
 \hline
 & \multicolumn{3}{|c|}{URE (full)} &\multicolumn{3}{|c|}{w/o using bootstrapping} \\
  \hline
  Dataset & U-Net & DCN & TransUNet & U-Net &DCN&  TransUNet \\
 \hline
   BUSI& 9.4($\pm$2.2)& 10.9($\pm$3.2) & 12.3($\pm$3.1) & 48.8($\pm$5.6) & 57.3($\pm$1.4)&  50.4($\pm$5.2) \\
    \hline
  ISIC& 8.5($\pm$3.0) &  8.8($\pm$3.6) &10.6($\pm$2.5)  & 54.3($\pm$4.5) &51.2($\pm$4.9) &48.6($\pm$5.7) \\
   \hline
\end{tabular}
\label{table:robustness}
\end{center}
\end{table*}

\subsection{Estimated Usable Regions on New Unseen Samples}
We examine the robustness of the URs on new unseen samples. For a given dataset, we randomly split the original test set into two halves, and use the 1st half to estimate UR (computing $\tau^*$) and the 2nd half to test whether the new samples satisfy the correctness requirement used in computing the $\tau^*$. We apply the above estimate-and-test procedure to a model 100 times, each time with a different random data split. We then report the frequency when the correctness of new samples violates the correctness requirement associated with the estimated UR. A low frequency indicates the estimated UR is robust to new samples. We repeat the whole procedure 20 times for reporting the mean and standard derivation for this frequency. Table \ref{table:robustness} (Left) shows that the estimated URs deliver fairly robust estimates, suggesting that they are informative in guiding model deployment. By comparing to the right part of Table \ref{table:robustness}, we validate the necessity of using bootstrapping for checking satisfaction of correctness in URE. 

%Checking correctness via comparing to the lower end of CI ensures a conservative sufficient estimate which yields significantly fewer violations when using the estimated UR on new samples. 

%\subsection{Statistics for More Conservative UR}
%As mentioned earlier, the statistics of the correctness scores used for summarizing the correctness scores in URE (in line 11 of Listing 1.1) are application-dependent and should be chosen by medical experts in practice. So far, ``Mean'' statistics are used in the above method description and experiments. In Table.., we show the effects on the UR size and the robustness of the UR for unseen samples when performing ``5 percentile'' for summarizing and testing correctness. ...
%\subsection{Utilities in Model Selection}
\noindent\textbf{Acknowledgement.} This work was supported in part by a grant from the Jiangsu Sci\&Tech Program (Y.Z.) and NSF Grant CCF-1617735 (D.Z.C.).

\section{Conclusions}

In this paper, we investigated the importance of measuring practical usability in evaluating medical AI segmentation models. We proposed CCRC for measuring the goodness of alignment between confidence estimates and correctness scores in rank, and developed a novel URE to quantitatively assess usable regions of predictions for AI models. Experiments validated that CCRC and URE are useful measures, bringing new insights for evaluating, comparing, and selecting models. 
\bibliographystyle{splncs04}
\bibliography{sample}
\end{document}